\documentclass[aps,reprint,showpacs,superscriptaddress]{revtex4-1} 
\usepackage{epsf}
\usepackage{graphicx}
\usepackage{sidecap}
\usepackage{amsmath}

\usepackage{dcolumn}   
\usepackage{bm}        
\usepackage{amssymb}   

\begin{document}

\title{Tunneling Tuned Spin Modulations in Ultrathin Topological Insulator Films}

\author{M.~Neupane}
\affiliation {Joseph Henry Laboratory and Department of Physics,
Princeton University, Princeton, New Jersey 08544, USA}

\author{A. Richardella}
\affiliation{Department of physics, The Pennsylvania State University, University Park, PA 16802}

\author{J. S\'anchez-Barriga}\affiliation {Helmholtz-Zentrum Berlin f\"ur Materialien und Energie, Elektronenspeicherring BESSY II, Albert-Einstein-Strasse 15, D-12489 Berlin, Germany}

\author{S.-Y.~Xu}
\affiliation {Joseph Henry Laboratory and Department of Physics,
Princeton University, Princeton, New Jersey 08544, USA}

\author{N.~Alidoust}
\affiliation {Joseph Henry Laboratory and Department of Physics,
Princeton University, Princeton, New Jersey 08544, USA}

\author{I.~Belopolski}
\affiliation {Joseph Henry Laboratory and Department of Physics,
Princeton University, Princeton, New Jersey 08544, USA}

\author{Chang~Liu}
\affiliation {Joseph Henry Laboratory and Department of Physics,
Princeton University, Princeton, New Jersey 08544, USA}

\author{G.~Bian}
\affiliation {Joseph Henry Laboratory and Department of Physics,
Princeton University, Princeton, New Jersey 08544, USA}

\author{D. M. Zhang}
\affiliation{Department of physics, The Pennsylvania State University, University Park, PA 16802}


\author{D. Marchenko}\affiliation {Helmholtz-Zentrum Berlin f\"ur Materialien und Energie, Elektronenspeicherring BESSY II, Albert-Einstein-Strasse 15, D-12489 Berlin, Germany}
\author{A. Varykhalov}
\affiliation{Helmholtz-Zentrum Berlin f\"ur Materialien und Energie, Elektronenspeicherring BESSY II, Albert-Einstein-Strasse 15, D-12489 Berlin, Germany}
\author{O. Rader}\affiliation {Helmholtz-Zentrum Berlin f\"ur Materialien und Energie, Elektronenspeicherring BESSY II, Albert-Einstein-Strasse 15, D-12489 Berlin, Germany}

\author{M. Leandersson}\affiliation {MAX-lab, P.O. Box 118, S-22100 Lund, Sweden}
\author{T. Balasubramanian}\affiliation {MAX-lab, P.O. Box 118, S-22100 Lund, Sweden}


\author{T.-R. Chang}
\affiliation{Department of Physics, National Tsing Hua University, Hsinchu 30013, Taiwan}

\author{H.-T. Jeng}
\affiliation{Department of Physics, National Tsing Hua University, Hsinchu 30013, Taiwan}
\affiliation{Institute of Physics, Academia Sinica, Taipei 11529, Taiwan}

\author{S.~Basak}
\affiliation {Department of Physics, Northeastern University,
Boston, Massachusetts 02115, USA}

\author{H.~Lin}
\affiliation {Department of Physics, Northeastern University,
Boston, Massachusetts 02115, USA}


\author{A.~Bansil}
\affiliation {Department of Physics, Northeastern University,
Boston, Massachusetts 02115, USA}

\author{N. Samarth}
\affiliation{Department of physics, The Pennsylvania State University, University Park, PA 16802}

\author{M.~Z.~Hasan}
\affiliation {Joseph Henry Laboratory and Department of Physics,
Princeton University, Princeton, New Jersey 08544, USA}

\date{\today}

\begin{abstract}
Quantitative understanding of the relationship between quantum tunneling and Fermi surface spin polarization is key to device design using topological insulator surface states. By using spin-resolved photoemission spectroscopy with $p-$polarized light in topological insulator Bi$_2$Se$_3$ thin films across the metal-to-insulator transition, we observe that for a given film thickness, the spin polarization is large for momenta far from the center of the surface Brillouin zone. In addition, the polarization decreases significantly with enhanced tunneling realized systematically in thin insulating films, whereas magnitude of the polarization saturates to the bulk limit faster at larger wavevectors in thicker metallic films. Our theoretical model calculations capture this delicate relationship between quantum tunneling and Fermi surface spin polarization. Our results suggest that the polarization current can be tuned to zero in thin insulating films forming the basis for a future spin-switch nano-device.
\end{abstract}


\maketitle

A three-dimensional topological insulator (TI) is a non-trivial phase of matter which acts as an electrical insulator in the bulk but can conduct a spin-polarized current on the surface \cite{Moore Nature insight, Fu Liang PRB topological invariants, David Nature BiSb, Matthew Nature physics BiSe, YLChen, SCZhang, David Nature tunable, QiKunXue, Sakamoto}. These surface states are characterized by a Dirac cone-like energy-momentum dispersion relation. The novel electronic structure of TIs can be manipulated to realize various novel quantum phenomena such as  spin-galvanic effects, dissipationless spin currents or neutral half-fermions for quantum information storage devices \cite{Galvanic effect, Qi Science Monopole, Essin PRL Magnetic, Yu Science QAH, Liang Fu PRL Superconductivity, Linder PRL Superconductivity}. The magnitude and wavevector dependence of the spin polarization of electrons and holes are among the most important  key ingredients in considerations for the design of any functional device. However, such developments have been severely hindered by residual bulk conductance in currently available materials which overwhelms the surface contribution. Additionally, scattering from the extrinsic bulk states leads to the reduction of spin polarization of the surface states. One promising route to minimize bulk conductance and thus improve effective spin polarization is to work with ultrathin films where the surface to volume ratio is significantly enhanced \cite{Linder, HZLu} and surface current can potentially dominate. On the other hand, in this limit the desired spin polarization of the surface states is kinematically reduced near the metal-to-insulator transition in the ultrathin films where the spin behavior is not known \cite{QiKunXue, Sakamoto}. 

To date, no systematic quantitative or even any spin-sensitive experimental study has been reported in the ultrathin limit across the metal-to-insulator transition despite the direct relevance of spins in ultrathin film limits for nano-device fabrication as well as the  potential discovery of novel topological phenomena.
We report a systematic spin-resolved angle-resolved photoemission spectroscopy (SR-ARPES) and spin-integrated ARPES measurements on ultrathin Bi$_2$Se$_3$ thin films using $p-$polarization of light for the first time. Our measurements reveal that the spin polarization is large for larger wavevectors, and the polarization magnitude increases with reduced tunneling, and its magnitude saturates to the bulk limit at a faster rate at large electron momenta. These unique spin features of ultrathin films, evidently distinct from the 3D TI, open up new possibilities for devices not possible with bulk topological insulators. 


Spectroscopic measurements were performed on large ultrathin Bi$_2$Se$_3$ films prepared by the  Molecular Beam Epitaxy (MBE) method on a GaAs(111)A substrates (Fig. 1a).
A compositional layout of the  film used in our measurements is shown in Fig. 1(b).
To protect the surface from contamination, about 40 nm Se capping is used on the top of ultrathin Bi$_2$Se$_3$ films. To expose the surface, the films were  transferred into the ARPES chamber and heated to 250 $^{\circ}$C at pressures lower than 1$\times$10$^{-9}$ torr about an hour which blow off the Se capping layer.  Our spin-resolved ARPES measurements were carried out in I3 beamline at Maxlab in Lund, Sweden, and
the UE112-PGM1 beamline with PHOENEXS chamber at BessyII in Berlin, Germany,
and the high-resolutions spin-integrated ARPES measurements were performed in beamlines 10.0.1 and 12.0.1 at the ALS. Fig. 1(c) shows the ARPES core level spectroscopy measurement of the unltrathin film before and after decapping of Se layer. Before decapping only Se peaks are visible while both Se and Bi peaks are observed after the decapping process, which proves that the Se capping works well in ultrathin film TI system. Thin films are characterized  by atomic force microscopy (AFM) (see Fig. 1(d)) and found that the root mean square (rms) surface roughness on these films is in the order of $\sim$ 0.2nm, which confirms the high quality of the films used in our measurements. The transport measurements of the Se capped ultrathin films result in carrier concentration, mobility and resistivity in the order of 1$\times$10$^{19}$cm$^{-3}$, 1270 cm$^{-2}$/Vs and 0.30 mOhm$\cdot$cm, respectively.



We used $p-$polarized light for our ARPES measurements. The photons come in the sample surface with angle of incident ($\theta$) 45 degree with the sample normal  and our samples are aligned along $\bar{\Gamma} - \bar{\text{K}}$ momentum-space cut (Fig. 2a) for spin-ARPES measurements.
The surface wavevector dependent  spin polarization is obtained using a Mott-polarimeter (Fig. 2(a)), which measures two orthogonal spin components of a photoemitted electron \cite{Berntsen}.  In the polarimeter, a gold foil was used as a scattering source to generate an asymmetry of high energy photoelectrons into different divergent spin states (see \cite{Suyang_spin}  and \cite{SOM} for details). Each orthogonal spin-polarization component is selected by the orientation of a scattering plane defined by the incident beam direction of the photoelectron on the gold foil and the orientation of two electron detectors mounted on each side of the foil. For this experiment, the detector configuration was set in a way that the two spin components correspond to the in-plane and out-of-plane directions of the (111) plane of the sample.

We present high-resolution spin-integrated ARPES data  and corresponding energy distribution curves (EDCs) along the high symmetry line $\bar{\Gamma} - \bar{\text{K}}$ for 1QL, 3QL, 4QL, 6QL and 7QL Bi$_2$Se$_3$ films in Figs. 2(c) and 2(d). 
As long as the thickness of the film is comparable to the decay length of the surface states into the bulk, there is a spatial overlap between the top and bottom surface states resulting in an energy gap at the time-reversal invariant point ($\bar\Gamma$ point).
As expected theoretically \cite{Linder, HZLu}, the energy gap decreases and eventually vanishes for sufficiently thick films, corresponding to the transition from a 2D gapped system (insulator) to a 3D gapless system (metal) \cite{QiKunXue, Sakamoto}.
In particular, the gapless dispersion relation observed in the 7QL film from ARPES measurement indicates that this thickness is above the quantum tunneling limit. 
Based on experimental observations, we present an illustration of the spin configuration for 3QL (insulator) and 7QL (metal) ultrathin Bi$_2$Se$_3$ films in Fig. 2(b).


We investigate the degree of spin polarization of the films as a function of electron wavevector for films of various thickness. To illustrate the wavevector dependent spin polarization, 3QL film is chosen ( see \cite{SOM} for detailed).
The SR-ARPES data for the wavevectors $\sim$ 0.05 \AA$^{-1}$ and 0.1 \AA$^{-1}$  are presented in Fig. 3. For each wavevector selected we present a spin-resolved energy distribution curve (EDC), which shows the relative intensity of photoelectrons with up and down spin polarizations (Figs. 3(a) and 3(c)).
For each such plot we associate a net spin polarization with the surface state of a given wavevector (Figs. 3(b) and 3(d)) and for 3QL film about 25 $\%$ and 15 $\%$ are estimated at momenta $\sim$ 0.1 \AA$^{-1}$  and $\sim$ 0.05 \AA$^{-1}$, respectively. 
 From these plots it is clearly observed that the net spin polarization decreases for smaller wavevectors (locations k $\sim$ 0.05 \AA$^{-1}$ in Fig. 3(d)).
The decrease can be understood as the presence of a tunneling gap in the ultrathin limit that effectively prevents the partner-switching behavior expected in the gapless topological surface states system \cite{HZLu}. The tunneling gap for ultrathin films can be seen in  the data (Fig. 2(c)).  
The reduction of the spin polarization and the existence of a tunneling gap in the data suggest that the bottom surface's right-handed contribution to the spin polarization must increase to effectively cancel the top surface's left-handed helical spin texture upon approaching smaller momenta values near $\bar{\Gamma}$.


Fig. 3(a) shows the spin-resolved EDCs  for 3QL, 4QL, 6QL and 60QL films while the corresponding net polarization is shown in Fig. 3(b) at $k$ = 0.1 \AA$^{-1}$. Analogous measurements are shown in Figs. 3(c) and 3(d) at $k$ = 0.05 \AA$^{-1}$ for 1QL, 3QL, 4QL, 6QL and 60QL  films. 
For thinner films, such as 3QL, there is a naturally larger contribution from the bottom surface (due to tunneling between top and bottom surfaces). This leads to a larger variation in spin polarization as a function of wavevector. For thicker films, this tunneling contribution decreases and the surface spin polarization becomes more uniform with varying wavevector. For instance, in the 60QL film, no measurable change in spin polarization is observed for the variation in wavevector magnitude. Our data suggest that the magnitude of polarization tends to reach the bulk limit faster at larger wavevectors.
 Experimental results are summarized in Fig. 4(a). Data further suggest that the relative contribution from the top surface systematically increases with film thickness.

First-principles theoretical modeling of the spin polarization behavior in thin films is presented in Figs. 4(b) and 4(c). 
In the calculations, symmetric slabs are used to simulate the thickness of  films. While a gapless spin-polarized Dirac cone is seen on the surface of a semi-infinite crystal of the topological insulator Bi$_2$Se$_3$, a gap is found to  open at the Dirac point for thin films due to a finite tunneling amplitude between the two sides of the slab surface in our calculations. The tunneling amplitude increases as the thickness decreases, causing  the gap to increase and
spin polarization to decrease in the gap region.

In the calculation we also consider the electron attenuation length ($\lambda$) due to the scattering processes since only electrons near the surface are able to reach the top of the surface and escape into the vacuum as in the measurement condition. Indeed, the spin polarization obtained by ARPES reflects the spin-texture of the states associated with top surface rather than the bottom surface.
The calculated spin expectation value for the electrons that can escape from the sample is  $\langle S \rangle_{\text{atom}} \times \exp ({-d_{\text{atom}}/\lambda}$), where $d_{\text{atom}}$ is the distance of an atom to the top surface, and the $\langle S \rangle_{\text{atom}}$ is the spin expectation value for each atom. The contribution from each atom is weighted by exp($-d_{\text{atom}}/\lambda$), which reduces the contribution from the bottom layer. Figs. 4(b) and 4(c) show the calculated results with $\lambda$ = 8\AA.

 It is important to note that the maximum spin polarization observed in the bulk limit is only about 40\% (Fig. 4(c)) whereas the original ideal theoretical limit is that of nearly 100\%  without considering any specific material system \cite{Kane}.
 In real TI materials, the strong spin-orbit interaction entangles the spin and orbital momenta of different atomic types, resulting in the reduction of spin polarization \cite{spin}. Specifically, the low-energy states in Bi$_2$Se$_3$ arise from $p$-orbitals of Bi (6$p^3$) and Se (4$p^4$), mostly $p_z$ levels of Bi and Se \cite{Matthew Nature physics BiSe, SCZhang}. The dominance of the $p_z$ orbitals in the topological surface states is further suggested by our circular dichroism (CD) measurements \cite{SOM}. The spin-orbit coupling mixes spin and orbital angular momenta while preserving the total angular momentum \cite{Matthew Nature physics BiSe, SCZhang}. The hybridization of orbitals in Bi and Se together with the entanglement of their spins contribute to the reduction of net spin polarization in real materials. 
Moreover, under the experimental geometry used in our measurement with $p-$polarized light (which is most sensitive to $p_z$ orbitals and most reflective of initial groundstate of the wavefunction), the penetration depth of the ARPES experiment (3-5 atomic layers maximum), the experimental observation of spin polarization is well agreed with recent theoretical calculations \cite{Damascelli, Park}.

Most importantly, our systematic  spin-spectroscopy results  suggest that  ultrathin films can serve as the basis for making qualitatively new devices, not possible with much-studied conventional 3D bulk topological insulators, despite the reduction of polarization magnitude.
For spintronics applications, our results suggest that ultrathin topological insulator films can be used to fabricate new types of nano-devices due to  the novel spin configurations and their systematic modulations possible in the ultrathin limit.
One such potential application implied by our spectroscopic results is that of a polarization current switch. 
Spin spectroscopic results suggest that it should be possible to control the polarization magnitude by varying a gate voltage across a high quality insulating thin film (Figs. 4(d) and 4(e), see \cite{SOM} for the chemical gating by K-deposition and NO$_2$ adsorption).
This effect, which cannot be readily realized in the highly-polarized states of conventional  3D bulk TIs (Fig. 4(e) right), allows ultrathin TI films to serve as the basis for functional nano-devices which can encode electrical signals in varying spin polarization magnitude or forms the physical basis for spin switch, among many other new application possibilities suggested by our observations of fundamental spin modulation behavior in ultrathin films.

\textit{Acknowledgements.} 
Sample growth and ARPES characterization are supported by US DARPA (N66001-11-1-4110).
The work at Princeton and Princeton-led synchrotron X-ray-based measurements and the related theory at Northeastern University are supported by the Office of Basic Energy Science, US Department of Energy (grants DE-FG-02-05ER462000, AC03-76SF00098 and DE-FG02-07ER46352).
M.Z.H. acknowledges visiting-scientist support from Lawrence
Berkeley National Laboratory and additional support from the
A. P. Sloan Foundation.
The spin-resolved and spin-integrated
photoemission measurements using synchrotron X-ray facilities are supported by
the Swedish Research Council, the Knut and Alice Wallenberg Foundation, the German Federal Ministry
of Education and Research, and the Basic Energy Sciences of the US Department of
Energy. Theoretical computations are supported by the US Department of Energy
(DE-FG02-07ER46352 and AC03-76SF00098) as well as the National Science Council and
Academia Sinica in Taiwan, and benefited from the allocation of supercomputer time at
NERSC and Northeastern University's Advanced Scientific Computation Center. 
We also thank
S-K. Mo and A. Fedorov for beamline assistance on spin-integrated photoemission
measurements (supported by DE-FG02-05ER46200) at Lawrence Berkeley National
Laboratory (The synchrotron facility is supported by the US DOE).


\begin{figure*}
\includegraphics[scale=0.17]{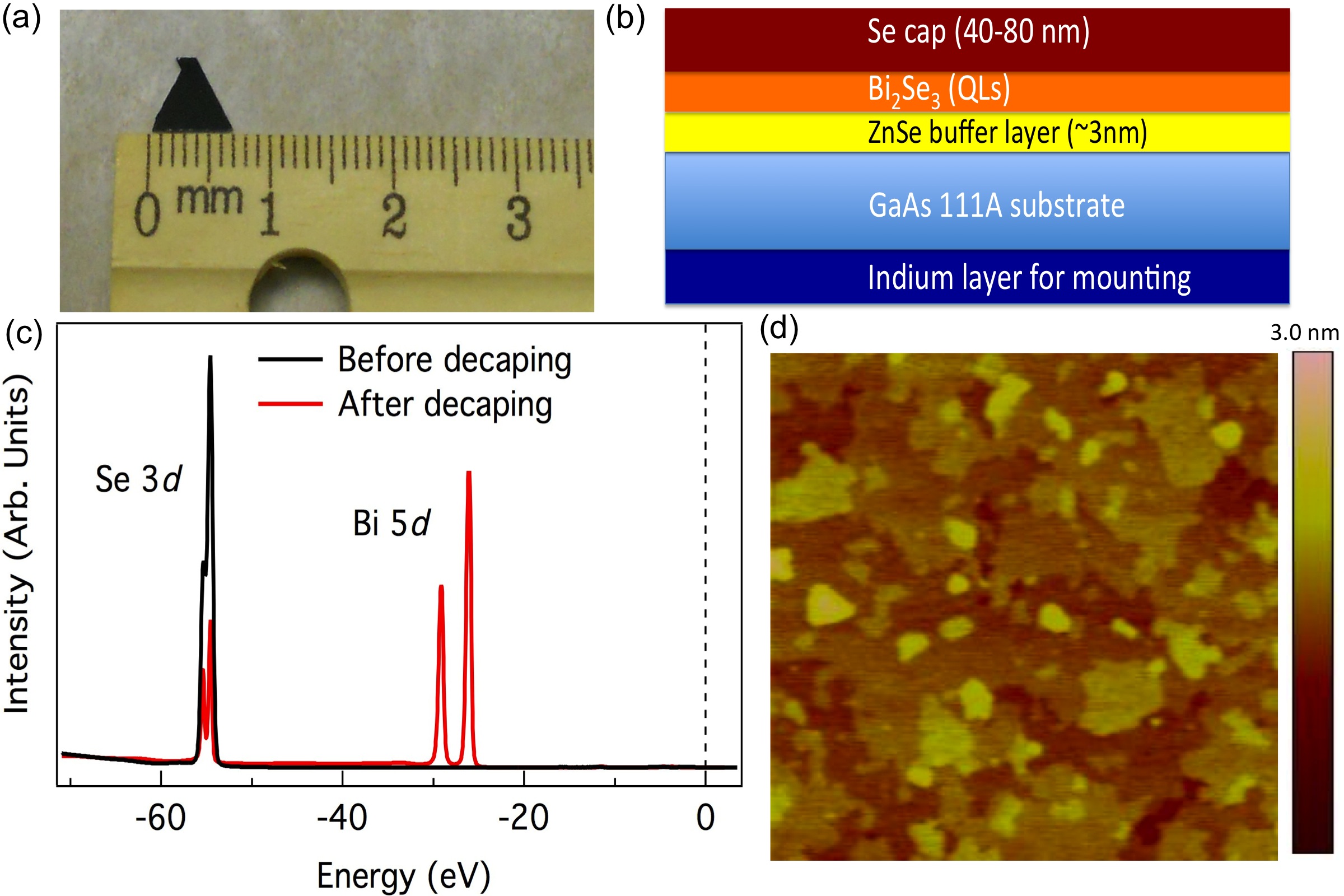}
\caption{Sample layout and characterization of ultrathin Bi$_2$Se$_3$.
(a) Photograph of a representative thin film sample used in SR-ARPES measurements. 
(b) Sample layout of ultrathin Bi$_2$Se$_3$ film, grown by MBE.
(c) Core level spectroscopy measurements on ultrathin MBE film before and after the decapping procedure. 
(d) AFM image of ultrathin Bi$_2$Se$_3$ film. }
\end{figure*}

\begin{figure*}
\centering
\includegraphics[scale=0.17]{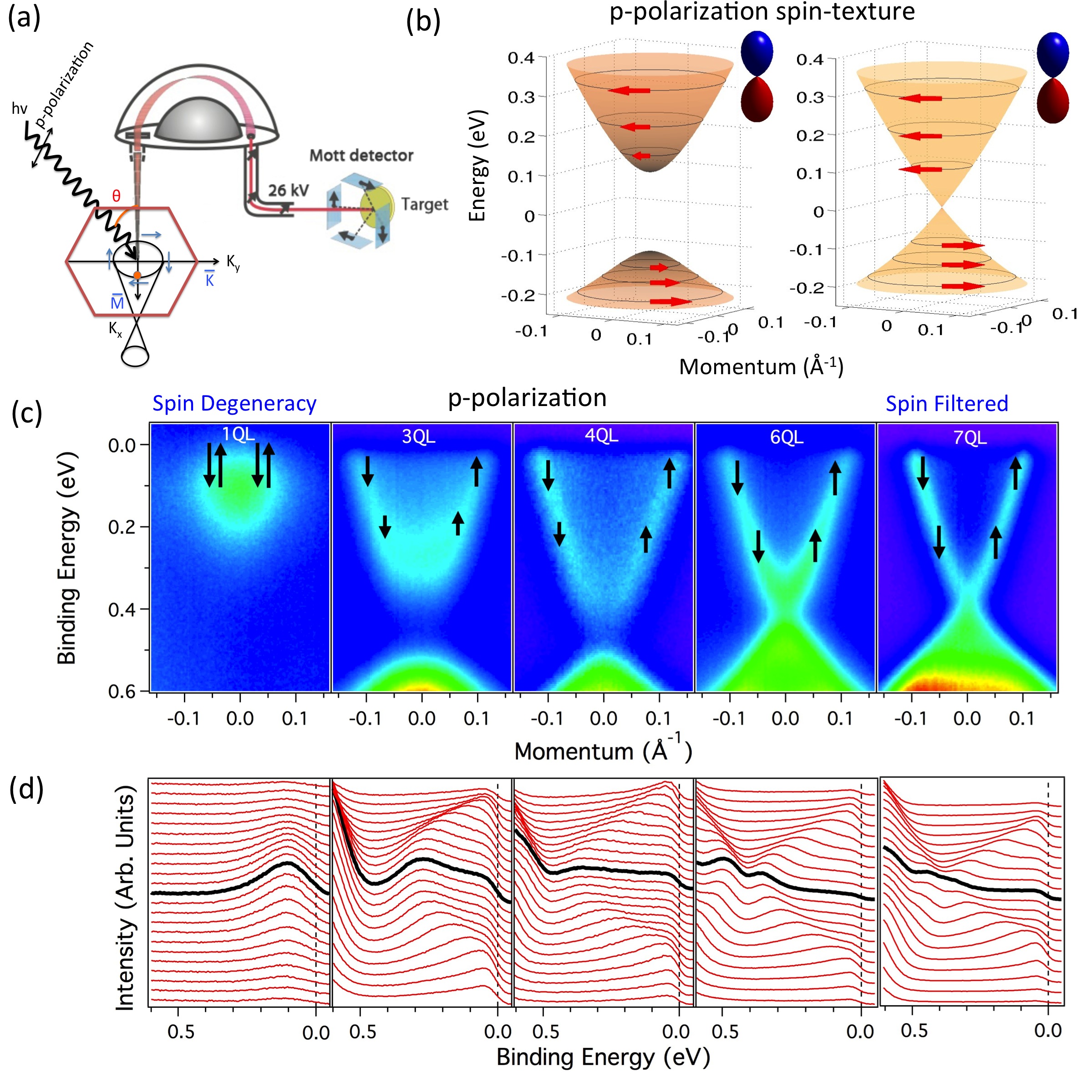}
\caption{Spin versus quantum tunneling in ultrathin Bi$_2$Se$_3$.
(a) Experimental geometry used in our measurements.
(b) Visualization of the contrasting spin configurations in 3QL (insulator) and 7QL (metal)  thin films. The dumbbell signs indicate that the current experimental geometry mainly probes the $p_z$ orbitals of Bi and Se.
(c) High-resolution ARPES measurements on ultrathin films of Bi$_2$Se$_3$:  $E-k$ band dispersion images for 1QL, 3QL, 4QL, 6QL and 7QL of  Bi$_2$Se$_3$ films taken near the $\bar{\Gamma}$ point along $\bar{\Gamma}-\bar{\text{K}}$ high-symmetry direction. The spin configuration is noted on the plots.
(d) The corresponding energy distribution curves (EDCs). The EDC through the $\bar{\Gamma}$ point is highlighted.}
\end{figure*}

\begin{figure*}
\centering
\includegraphics[scale=0.5]{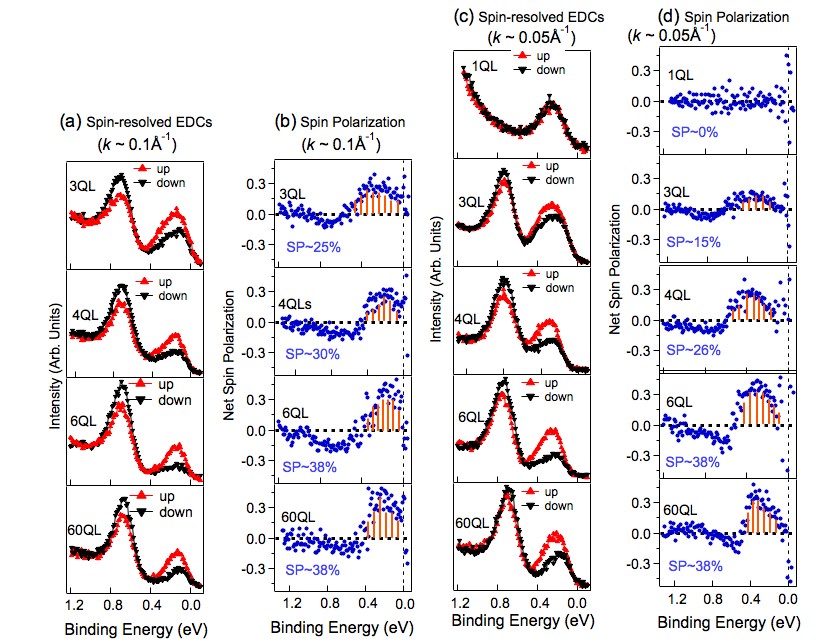}
\caption{Thickness dependent quantum tunneling and evolution of spin configuration.
(a) Spin-resolved EDCs and (b) net tangential spin polarizations for 3QL, 4QL, 6QL and 60QL ultrathin Bi$_2$Se$_3$ films at \textit{k} $\sim$ 0.1 \AA$^{-1}$. (c-d), same as (a) and (b) for 1QL, 3QL, 4QL, 6QL and 60QL ultrathin Bi$_2$Se$_3$ films at \textit{k} $\sim$ 0.05 \AA$^{-1}$. 
The red (black) curves show tangentially up (down) spin-resolved EDCs. The magnitude of each net spin polarization is also noted in (b) and (d). 
The vertical red lines in  (b) and (d) are guides to the eye, indicating a non-zero area under the spin polarization curve.
We note that 1 quintuple layer (QL) is equivalent to $\sim$ 1 nm.}
\end{figure*}

\begin{figure*}
\includegraphics[scale=0.45]{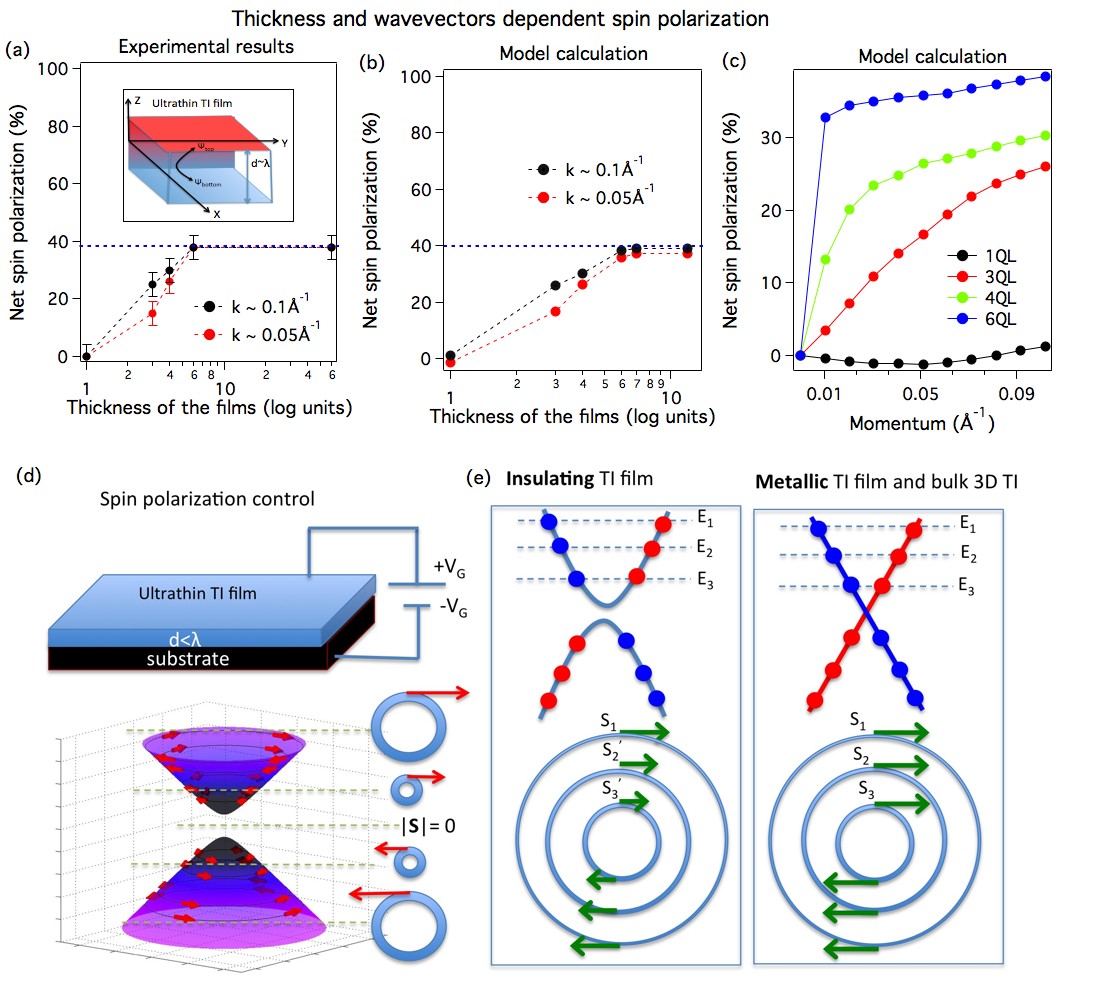}
\caption{ Experimental versus theoretical spin polarization.
(a) Experimentally-observed net spin polarization as a function of thin film thickness for $k \sim$ 0.05  \AA$^{-1}$ and 0.1 \AA$^{-1}$. The inset shows a schematic view of quantum tunneling in ultrathin TI  films.
(b) Calculation results of spin polarization versus film thickness at two momentum points $k \sim$ 0.05  \AA$^{-1}$ and 0.1 \AA$^{-1}$. 
(c) Results of a calculation of the net spin polarization as a function of wavevector for ultrathin films of thickness 1QL, 3QL, 4QL, and 6QL.
Dashed lines in (a,b) and solid lines in (c) between the dots serve as guides to the eye. Error bars in (a) represent experimental uncertainties in determining the spin polarization. 
Schematic of  (d) a gate controlled spin polarization current switch device and
(e) momentum dependent spin configuration in ultrathin (insulating) film and thicker (metallic) film. }
\end{figure*}

\end{document}